\documentclass[aps,prb,reprint,superscriptaddress]{revtex4-2}
\usepackage{amsmath}
\usepackage{graphicx}
\usepackage{bbm}
\usepackage[colorlinks,linkcolor=blue,anchorcolor=blue,urlcolor=blue,citecolor=blue]{hyperref}
\usepackage{graphicx}
\usepackage{dcolumn}
\usepackage{bm}
\usepackage[utf8]{inputenc}
\usepackage[T1]{fontenc}
\usepackage{etoolbox}
\usepackage{amssymb}
\usepackage{yhmath}
\makeatletter

\begin{document}

\title{Degenerate Topological Edge States in Multimer Chains}

\author{Jun Li}
\email[]{jli\_phys@tongji.edu.cn}
\affiliation{MOE Key Laboratory of Advanced Micro-Structured Materials, School of Physics Science 
	and Engineering, Tongji University, Shanghai, 200092, China}
\affiliation{Department of Physics and Astronomy, University of Manitoba, Winnipeg R3T 2N2, Canada}

\author{Yaping Yang}
\email[]{yang\_yaping@tongji.edu.cn}
\affiliation{MOE Key Laboratory of Advanced Micro-Structured Materials, School of Physics Science 
	and Engineering, Tongji University, Shanghai, 200092, China}

\author{C.-M. Hu}
\email[]{hu@physics.umanitoba.ca}
\affiliation{Department of Physics and Astronomy, University of Manitoba, Winnipeg R3T 2N2, Canada}

\date{\today}

\begin{abstract}
We propose and experimentally realize a class of quasi-one-dimensional topological lattices whose 
unit cells are constructed by coupled multiple identical resonators, with uniform hopping and inversion 
symmetry. In the presence of coupling-path-induced effective zero hopping within the unit cells, the 
systems are characterized by complete multimerization with degenerate $-1$ energy edge states for 
open boundary condition. Su-Schrieffer-Heeger subspaces with fully dimerized limits corresponding to 
pairs of nontrivial flat bands are derived from the Hilbert spaces. In particular, topological bound states 
in the continuum (BICs) are inherently present in even multimer chains, manifested by embedding the 
topological bound states into a continuous band assured by bulk-boundary correspondence. Moreover, 
we experimentally demonstrate the degenerate topological edge states and topological BICs in 
inductor-capacitor circuits.

\end{abstract}


\maketitle

Topological phases of matter transcend the paradigm of Ginzburg-Landau theory in condensed 
matter physics, with absence of any symmetry breaking but derived from geometry, and have 
attracted extensive investigation in various fields over the past few decades 
\cite{PhysRevLett.45.494, PhysRevLett.49.405,PhysRevLett.61.2015,RevModPhys.82.3045, 
	RevModPhys.83.1057,RevModPhys.88.021004, RevModPhys.88.035005, RevModPhys.89.041004, 
	RevModPhys.93.015005}. Topological phases are defined by the global wavefunctions of the 
dispersion bands that pervade the entire system rather than local orbitals, so that they are 
particularly robust to local perturbations such as defects and impurities. In essence, Band 
structure is the sufficient condition for the existence of topological phases. Since the first 
discovery of topological phases in quantum electronic systems \cite{PhysRevLett.45.494,
	PhysRevLett.49.405}, novel and exotic topological properties have been developed in diverse 
platforms with their own unique advantages such as optics \cite{Lu2014, RevModPhys.91.015006}, 
acoustics \cite{PhysRevLett.114.114301, nphys3228}, mechanics \cite{Huber2016} and electric 
circuit \cite{PhysRevLett.114.173902, PhysRevX.5.021031, Lee2018} in classical regimes and 
ultra-cold atoms \cite{RevModPhys.91.015005, PhysRevLett.119.023603}, trapped-ions 
\cite{nsr/nwy142, s41586-022-04853-4} and Fock-state lattices \cite{nsr/nwaa196, science.ade6219} 
in quantum regimes.

One-dimensional (1D) topological phases bring some new insights because of their manipulability 
and experimental accessibility. The Su-Schrieffer-Heeger (SSH) model of polyacetilene
\cite{PhysRevLett.42.1698, RevModPhys.60.781}, as a starting point for 1D topological models based 
on tight-binding approximation, is a dimerized chain by having two different alternating hopping 
amplitudes between nearest-neighboring lattice site hosts. Recently, in the context of 
SSH chains, a variety of extended configurations with new physics and phenomena has been 
proposed especially non-Hermitian topology \cite{RevModPhys.93.015005, PhysRevX.9.041015}. 
On the one hand, special inconsistent inter-site interactions, like periodically modulated hopping, 
nonreciprocal hopping, environment-induced coupling and multisite coupling, have been 
introduced to raise a plethora of distinct topological phenomena including but not limited to the 
non-Hermitian skin effect \cite{PhysRevLett.116.133903, PhysRevLett.121.086803, 
	PhysRevLett.123.246801, PhysRevLett.124.086801, s41567-020-0922-9}, non-Hermitian real 
spectra \cite{PhysRevB.105.L100102}, dissipative and Floquet topological phase transition \cite{PhysRevLett.127.250402,Fan:22,PhysRevB.106.224306} and trimer topological phases
\cite{PhysRevLett.46.738,PhysRevA.99.013833,s41598-021-92390-x,Yan:23}. On the other hand, 
with respect to on-site potentials, the introduction of on-site gain and loss not only provides a 
pointcut to combine the non-Hermitianity and topological phases for widening topological family 
\cite{nmat4811, PhysRevLett.123.165701, PhysRevLett.126.215302}, but also can drive topologically 
trivial systems and induce topological phase transitions solely by deliberate design 
\cite{PhysRevLett.121.213902, PhysRevLett.123.073601, PhysRevB.101.180303, 
	PhysRevLett.124.236403, PhysRevResearch.4.023009}.

In this letter, we present a quasi-one-dimensional (quasi-1D) tight-binding configuration without 
any staggered hopping and on-site potentials. We consider unit cells of multiple identical resonators 
with uniform coupling between every two sites and the same strength as the inter-cell coupling, i.e., 
only one kind of coupling strength and resonators in the whole chain. The system then forms 
complete multimer due to the zero effective intracell hopping induced by special coupling paths. 
Conceivably, considering the bulk-boundary correspondence (BBC), degenerate topological edge 
states with full localization exist in finite systems \cite{2016short}. Interesting, with the increase of 
resonators in the unit cell, pairs of non-trivial flat bands appear at two fixed frequencies which 
correspond to fully dimerized subspaces derived from the Hilbert spaces. Moreover, for even 
multimer chains, topological bound states in the continuum (BICs)  \cite{PhysRevLett.118.166803}
naturally form via the bandgap of nontrivial flat bands just covered by a trivial band. We experimentally 
implement the idea by using AC circuits consisting of uniform capacitors and inductors.

\begin{figure}
	\includegraphics[width=0.8\linewidth]{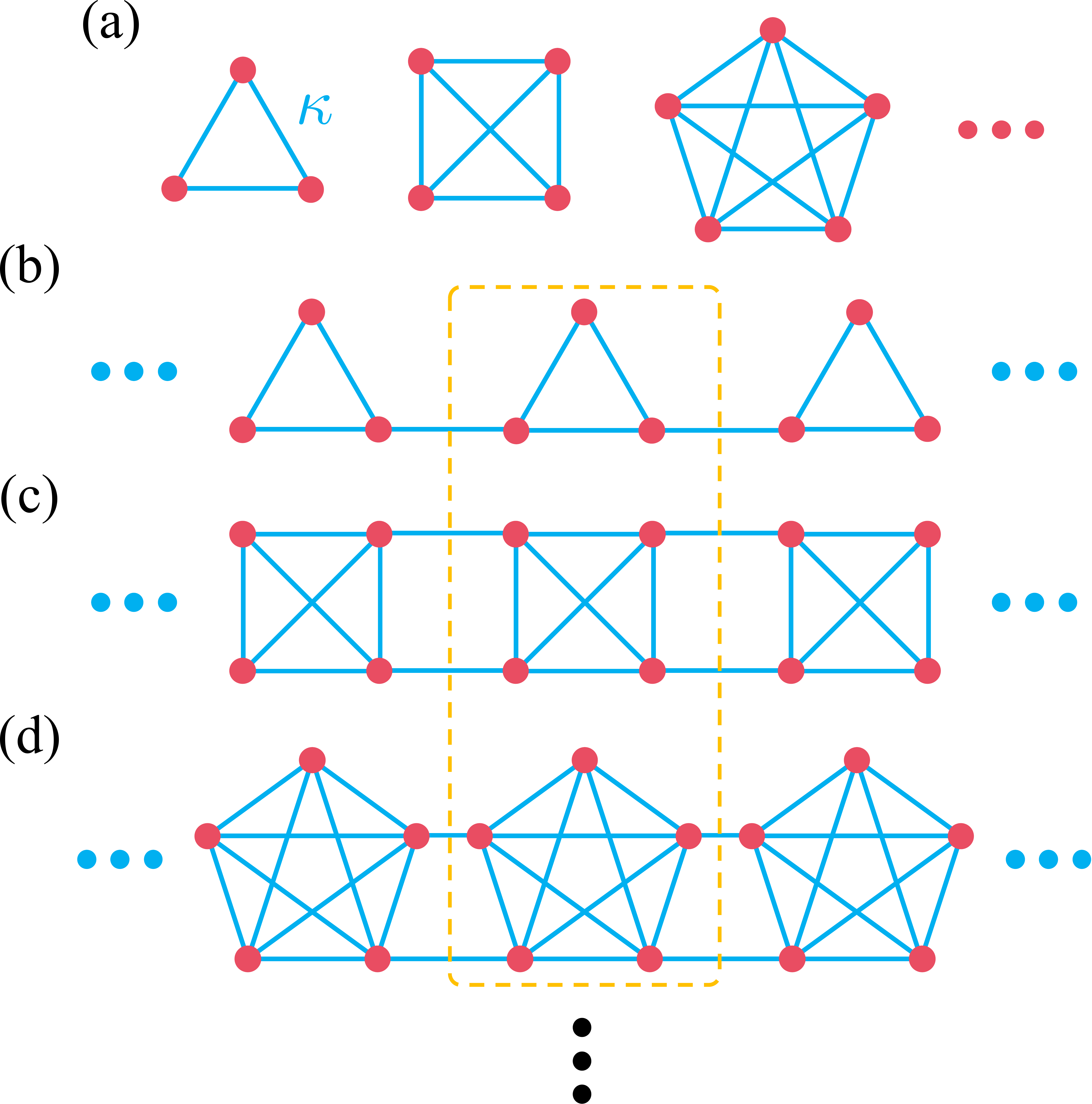}%
	\caption{\label{Fig.1} Theoretical tight-binding hopping model. (a) Schematic of $n$ $(n=3, 4, 
		5 \cdots)$ resonators coupled with each others with uniform hopping amplitude $\kappa$. 
		(b-d) Bulk model with $n$ sites per unit cell, with uniform hoppings, unit cells framed in yellow 
		dashed box.}
\end{figure}

We start by considering a tight-binding system consisting of $n$ $(n \ge 3)$ identical resonators 
coupled to each other with the same hopping amplitude $\kappa$, as shown in Fig. \ref{Fig.1} (a).  
Here, we consider an Hermitian system that the intrinsic and coupling losses of all the resonators 
are ignored and $\kappa$ is sufficiently small compared to the frequency of resonators 
$\omega_{0}$. the systems can be represented  by the Hamiltonian

\begin{equation}\label{eq:1}
	H_{n}=\left(\begin{array}{cccc}
		0          & \kappa & \kappa & \cdots \\
		\kappa & 0          & \kappa & \ddots \\
		\kappa & \kappa & 0          & \ddots  \\
		\vdots  & \ddots  & \ddots & \ddots
	\end{array}\right)_{n\times n},
\end{equation}
characterized that the diagonal elements of the matrix are zero and all the others are $\kappa$.
 Interestingly, there are always degenerate states with a fixed frequency independent of $n$ in the 
 system. Specifically, with reference to $\omega_{0}$, one of its eigenvalues is $\lambda_{n}=(n-1)
 \kappa$ with the normalized eigenvectors $\left|\psi_{n}\right\rangle=(1/\sqrt{n}, 1/\sqrt{n},\cdots,
 1/\sqrt{n})'$ while the others are $\lambda_{i}=-\kappa$ where $i=1,2,\cdots,n-1$ with the 
 corresponding eigenvectors $\left|\psi_{i}\right\rangle = (0,\cdots,0,1/\sqrt{2})'$ where the $i$th 
 element $\left|\psi_{i}\right\rangle _i=1/\sqrt{2}$. In terms of the splitting of eigenvalues, the local 
 effective coupling between the degenerate modes can be seen as zero to some extent.

With this supposition, as shown in Figs \ref{Fig.1} (b-d), we design a class of quasi-1D 
lattices with the above coupling multiple resonators as their unit cell. The unit cells are coupled 
to their nearest neighbor through $[n/2]$ independent coupling channels with the same hopping 
amplitudes $\kappa$. Considering the zero effective intracell hopping, we can expect that our 
chains are topologically nontrivial with complete multimerization. In bulk momentum space, the 
Bloch Hamiltonian of the chain can be written as
\begin{equation} \label{eq:2}
	H_{n}(k)=\left(\begin{array}{cccc}
		0                                     & \kappa & \cdots  & \kappa+\kappa e^{-ika} \\
		\kappa                            & 0          & \adots    & \kappa\\
		\vdots                             & \adots  & \ddots & \vdots\\
		\kappa+\kappa e^{ika} & \kappa & \cdots  & 0
	\end{array}\right)_{n\times n}
\end{equation}
where $a$ is the lattice constant between the units and $k$ is the Bloch wave number. The Bloch 
Hamiltonian shows that the Bloch term only exists in all anti-diagonal elements, provided that the 
diagonal term remains zero. Obviously, the Hamiltonian displays inversion ($\mathcal{I}$) symmetry, 
i.e., $\mathcal{I} H_{n}(k) \mathcal{I}^{-1}=H_{n}(-k)$. For clarity, the system is classified into two 
patterns via $n$ is odd or even in the following analysis. We find the analytical solutions of its energy 
spectra expressed as 
\begin{equation} \label{eq:3}
	\omega_{n, o}(k)=\left(\begin{array}{cccc}
		-2 \\
		\vdots \\
		n/3-1-(A_{+}+i\sqrt{3}A_{-})/2 \\
		n/3-1-(A_{+}-i\sqrt{3}A_{-})/2 \\
		0 \\
		\vdots \\
		n/3-1+A_{+} \\
	\end{array}\right)\kappa
\end{equation}
when $n$ is odd where $A_{\pm}=C/B \pm B$, $B = \left[\sqrt{D^2-C^3}+D\right]^{1/3}$, $C = (n-1) 
\cos {ka}/3+(n^2+3)/9$  and $D=(n^2-n)\cos{ka}/6+(n/3)^3+(n-3)/6$. Surprisingly, there are 
$n-3$ flat bands equally divided at $\omega_{n}=0$ and $\omega_{n}=-2\kappa$. And only two 
bandgaps with the width $G_1=\sqrt{(n-2)^2+1}-\sqrt{(n-1)^2-1}+1$ and $G_2=\sqrt{(n-2)^2+1}+n-2$
exist in the energy spectra. For the case where $n$ is even, the eigenfrequency is given by
\begin{equation} \label{eq:4}
	\omega_{n, e}(k)=\left(\begin{array}{cccc}
		-2 \\
		\vdots\\
		n/2-1-\sqrt{n^2/4+1+n\cos{ka}} \\
		0 \\
		\vdots \\
		n/2-1+\sqrt{(n^2/4+1+n\cos{ka}}
	\end{array}\right)\kappa,
\end{equation}
characterized by having $n/2-1$ flat bands at $\omega_{n}=0$ and $\omega_{n}=-2\kappa$ 
respectively. The ($n/2$)th band and the top ($n$th) band are symmetric with respect to 
$(n/2-1)\kappa$. It is noteworthy that as the parameter $n$ varies, there consistently exists one 
bandgap with a width of $G=n-2$ due to the lower non-flat band precisely overlapping the bandgap 
of the flat bands. In both cases, the band structure of the bulk Hamiltonian is not symmetric 
around zero, indicating that our chain is chiral symmetry broken. In detail, except for the top 
band that exceeds zero, the other bands are always distributed between $-2\kappa$ and zero.

\begin{figure*}
	\includegraphics[width=0.75\linewidth]{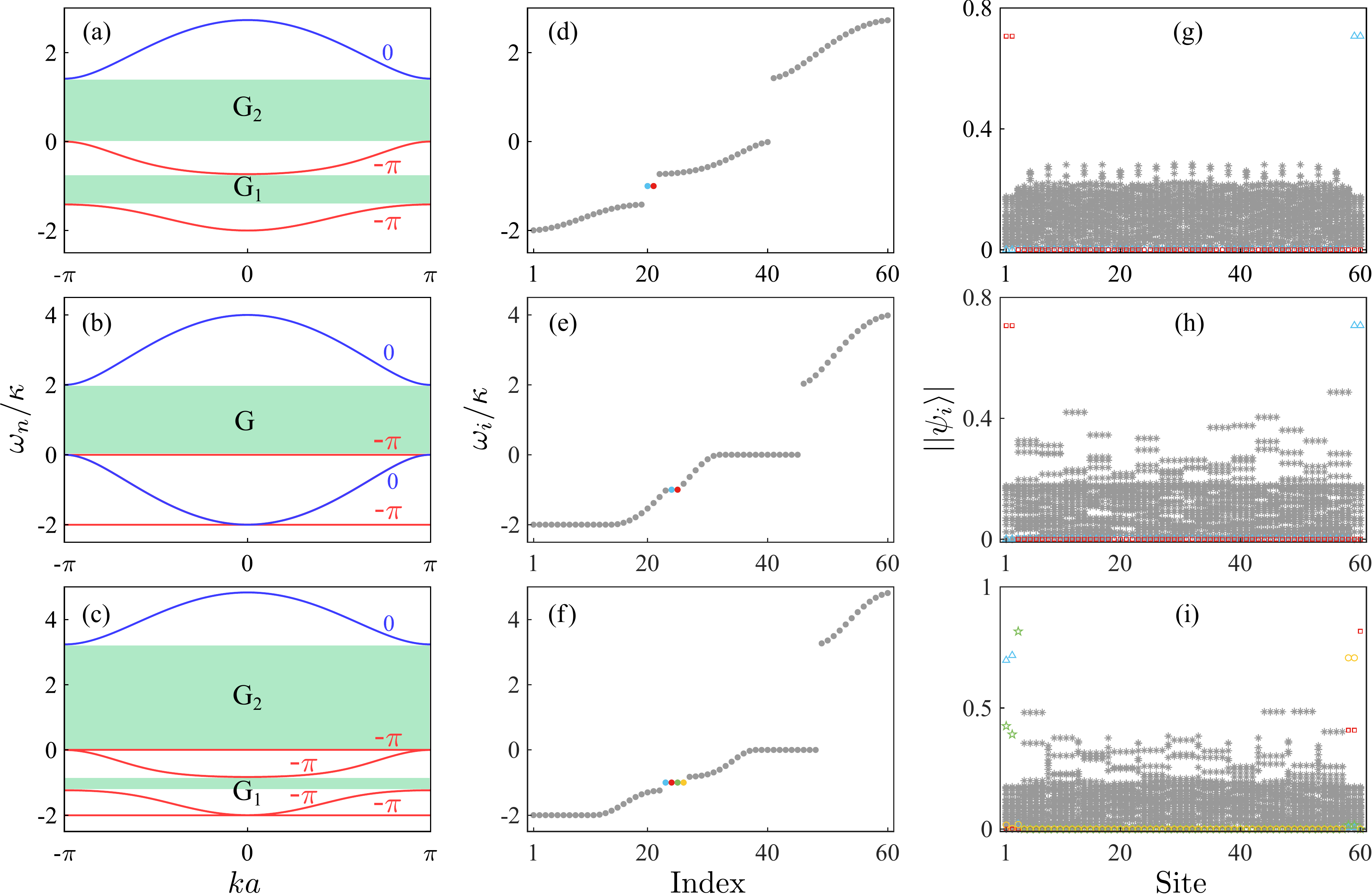}%
	\caption{\label{Fig.2}Topological edge states of multimer chains. (a-c) The normalized  band 
		structures with quantized Zak phases and (d-f) sorted eigenvalues of topological finite chains 
		(composed of 60 resonators) with (g-i) corresponding representative wave functions for (a, d, g) 
		$n=3$, (b, e, h) $n=4$ and (c, f, i) $n=5$, respectively. Zak phases for the bands labeled by red 
		are $-\pi$ and 0 by blue in (a-c). The edge and bulk states in (d-f) with the corresponding 
		intensity distributions in (g-i) are represented by color and gray, respectively. Particularly, the 
		wavefunction distributions in panels (h) and (i) represent individual instances of potential 
		numerical solutions for $n=4$ and $n=5$, respectively.}
\end{figure*}

Figures \ref{Fig.2} (a-c) show the band structures in first Brillouin zone for different $n$. Here, in the 
presence of inversion symmetry, we introduce the Zak phase, defined as $Z_{j} = -i \int_{-\pi / a}
^{\pi / a} \left\langle \psi_{k, j}\left|\partial_{k}\right| \psi_{k, j} \right\rangle d k$, to characterize the 
topology of our 1D multimer system where $j$ specifies the occupied band index with corresponding 
Bloch wave functions $\left|\psi_{k, j}\right\rangle$ \cite{PhysRevLett.62.2747}. We can obtain 
nonzero quantized Zak phases of bands for various $n$, indicating the topological nontriviality of our 
chains. Particularly, the top band possesses a Zak phase of zero while the flat bands always have Zak 
phases of $-\pi$. We can block-diagonalize $H_{n}(k)$ by unitary transformation $\mathcal{U}_{n}^{-1} 
H_{n}(k) \mathcal{U}_{n} = H_{n}^{BD}(k)$ to show the separation between flat and nonflat bands
 clearly. For the simplest case with $n=4$, the unitary matrix and the block-diagonal Hamiltonian are
\begin{equation}\label{eq:5}
	\mathcal{U}_4=\frac{1}{\sqrt{2}}\left(\begin{array}{cccc}
		1 & 0 & 1   & 0 \\
		1 & 0 & -1 & 0 \\
		0 & 1 & 0  & -1\\
		0 & 1 & 0  & 1
	\end{array}\right),
\end{equation}
\begin{equation}\label{eq:6}
	H_{4}^{BD}(k)=\left(\begin{array}{cccc}
		1                & 2+ e^{-ika} & 0            & 0 \\
		2+e^{ika} & 1                   & 0            &0 \\
		0               & 0                  & -1           & e^{-ika} \\
		0               & 0                  & e^{ika}  & -1
	\end{array}\right)\kappa,
\end{equation}
respectively. As expected, the both $2\times2$ blocks have the same form as the SSH Hamiltonian 
where the upper one is topologically trivial corresponding to the blue bands and the lower block is 
topologically nontrivial with complete dimerization corresponding to the flat bands. More generally, 
the bulk Hamiltonians for larger $n$ can be divided into a topological trivial dimerized subspace and 
$n/2-1$ nontrivial fully dimerized subspaces by the unitary transformation when $n$ is even. Moreover, 
the lower trival band for even chain always spans the bandgap between the flat bands by meeting 
the upper and lower flat bands at the boundary and center of the first Brillouin zone, respectively.
Similarly, for odd $n$, we can get the block-diagonal Hamiltonian composed of a $3\times3$ block 
and $(n-3)/2$ same nontrivial $2\times2$ blocks (see Supplemental Material for details \cite{sm}) 
where the $3\times3$ subspace owns two nontrivial lower bands in contact with the upper and lower 
flat bands independently. 

Considering the BBC, under the open boundary condition, we show the normalized eigenvalue spectra 
$\omega_i/\kappa$ of finite multimer chains with 60 resonators in Figs. \ref{Fig.2} (d-f) and the 
corresponding wave functions $\left|\psi_{i}\right\rangle$ in Figs. \ref{Fig.2} (g-i) for $n=3$, 4 and 5.  
Mathematically, the wavefunction solution is not unique for the finite-size chains with $n>3$, owning 
to the fact that the rank of the Hamiltonian matrix is smaller than the matrix dimension. The 
wavefunction distributions depicted in Figs. \ref{Fig.2} (h) and \ref{Fig.2} (i) exemplify potential 
numerical solutions for $n=4$ and $n=5$, respectively. Correspondingly, 
there are pairs of degenerate edge states marked by the colored dots at the exact detuning $-\kappa$ 
with the same number as the nontrivial bands. The topological edge modes of odd multimer chain sit in 
the lower bandgap $G_{1}$. Remarkably, the topological edge states are clearly embedded in the 
continuous spectrum of the lower nontrival band and are the so-called topological BICs 
\cite{PhysRevLett.118.166803} when $n$ is even. Because of the complete multimerization, the wave 
functions of exact $-1$ energy edge modes are absolutely localized at the two boundary cells without 
any distribution in the bulk, while the bulk wave functions are diffused throughout the whole chains.

\begin{figure}
	\includegraphics[width=\linewidth]{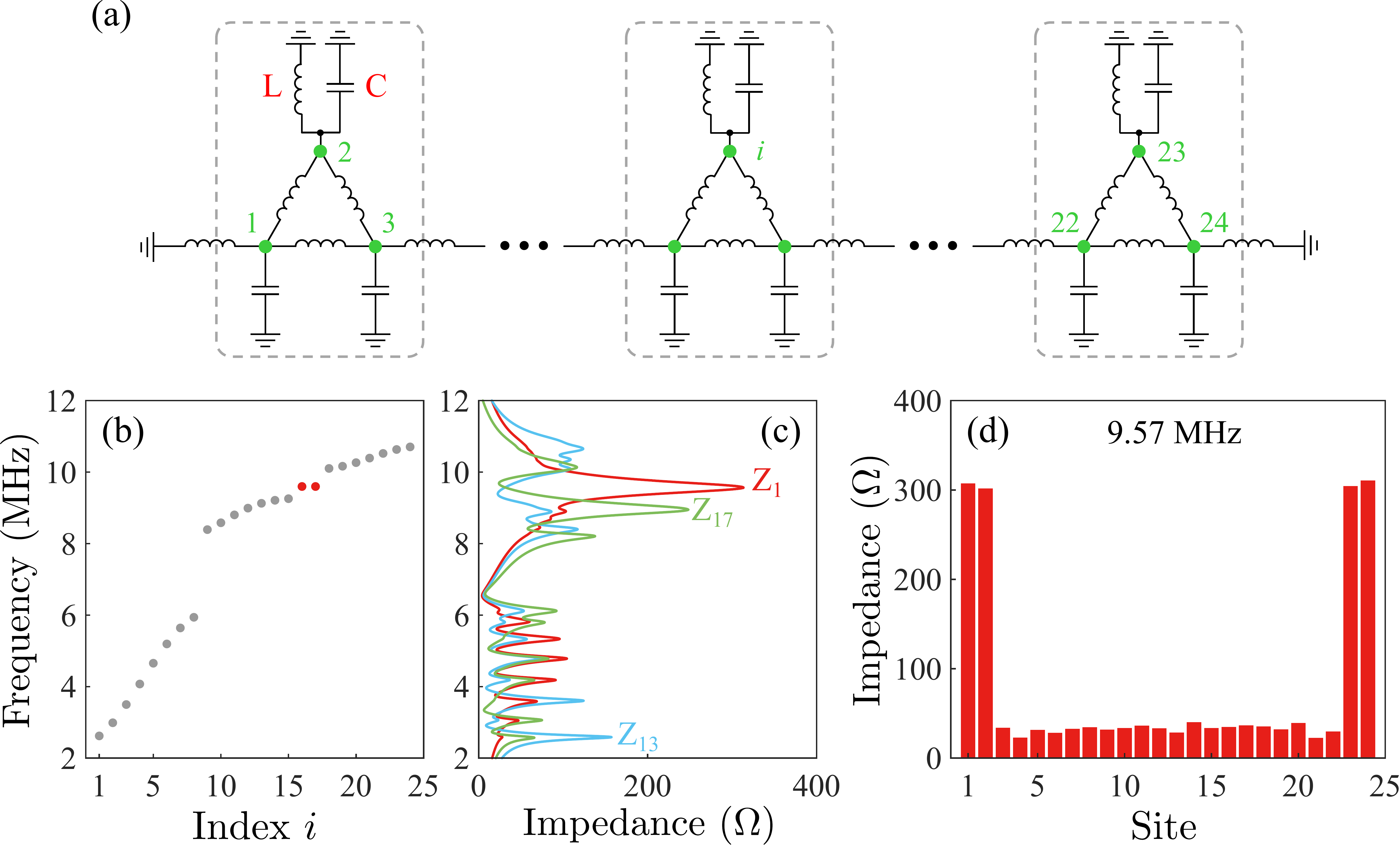}%
	\caption{\label{Fig.3} Observation of topological edge states in trimer chain. (a) Circuit diagram of 
		the finite experimental trimer chain, unit cells consists of three capacitors $C$ with identical 
		inductors $L$ between every two capacitors framed in grey dashed boxes. (b) Calculated 
		admittance eigenspectrum of the LC circuit for $C=1$ nF and $L=1.1$ $\mu$H. (c) Measured 
		impedances between the nodes ($|Z_{1}|$, $|Z_{13}|$ and $|Z_{17}|$) and ground vs the frequency 
		of circuit. (d) Location distribution of impedance at the frequency $f=9.57$ MHz.}
\end{figure} 

\begin{figure}
	\includegraphics[width=\linewidth]{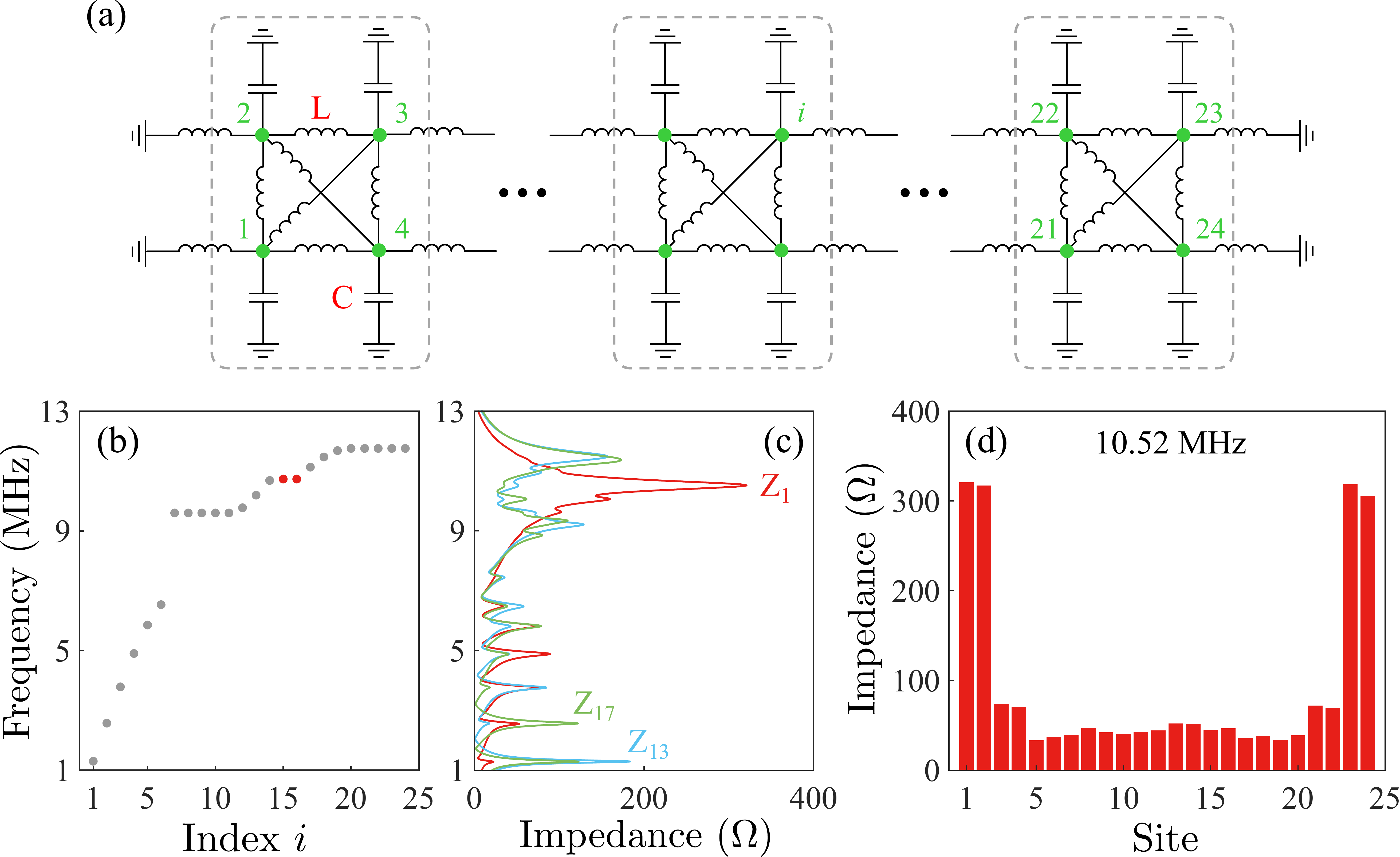}%
	\caption{\label{Fig.4} Observation of topological BICs in tetramer chain. (a)Circuit diagram blueprint 
		with unit cells framed in grey dashed boxes. (b) Calculated admittance eigenvalues of the tetramer 
		LC circuit for $C=1$ nF and $L=1.1$ $\mu$H. (c) Frequency scan of measured impedances for 
		representative edge ($|Z_{1}|$) and bulk ($|Z_{13}|$ and $|Z_{17}|$) nodes. (d) Impedance 
		distribution of topological edge mode at the frequency $f=10.52$ MHz.}
\end{figure}

We employ periodic inductor-capacitor (LC) circuits featuring flexible hopping channels to 
experimentally observe the tight-binding modes. Here, the lattice nodes are capacitively coupled to 
ground and inductively coupled to each other. The multimer chains can be represented by the 
admittance matrix $J_{\omega}$ (also termed circuit Laplacian) \cite{Lee2018, 
	PhysRevLett.126.215302}. The voltage response $\boldsymbol{V} (\omega)$ of the nodes to a input 
current $\boldsymbol{I}(\omega)$ at frequency $\omega$ follows Kirchhoff’s law: 
$\boldsymbol{I}(\omega)=J(\omega) \boldsymbol{V}(\omega)$ where the vectors 
$\boldsymbol{I}(\omega)=\left[I_1, I_{2},\cdots,I_{s}\right]'$ and $\boldsymbol {V}(\omega)=\left[V_{1}, 
V_{2},\cdots,V_{s}\right]'$ for $s$ nodes circuit. For our uniform hopping chains, using the same size 
capacitors $C$ and inductors $L$, we have the circuit Laplacian
\begin{equation}\label{eq:7}
	J(\omega)=\frac{1}{i\omega} \left[\left(\frac{n}{L}-\omega^2 C\right) \mathbb{I} +H\right]
\end{equation}
with
\begin{equation}\label{eq:8}
	H=\left(\begin{array}{cccc}
		0          & -1/L      & -1/L      & \cdots \\
		-1/L      & 0          & -1/L      & \cdots \\
		-1/L      & -1/L      & 0          & \cdots  \\
		\vdots  & \vdots  & \vdots  & \ddots
	\end{array}\right)_{s\times s},
\end{equation}
where $\mathbb{I}$ is the $s\times s$ unit matrix and $H$ can represent our theoretical model 
accurately with the hopping amplitude $-1/L$. We construct periodic LC circuits with 24 nodes for 
trimer and tetramer configurations as shown in Fig. \ref{Fig.3} (a) and Fig. \ref{Fig.4} (a), respectively. 
By solving the eigenvalues $E_{i}$ of $H$ numerically, we can obtain the general admittance 
eigenspectra dispersion as $f_i=\sqrt{(n/L+E_{i})/C}\big/(2\pi)$ with degenerate edge states labeled by 
red shown in Figs. \ref{Fig.3} (b) and \ref{Fig.4} (b). Note that the inverted nonlinear spectra is due to 
the negative frequency-dependent hopping amplitude of inductive coupling. 

In the experiment implementation, we choose the circuit components: $C=1$ nF with $\pm1\%$ tolerance 
and $L=1.1$ $\mu$H with $\pm5\%$ deviation. Details of the sample fabrication and impedance 
measurements are provided in the Supplemental Material \cite{sm}. Measured impedances of nodes 1, 13, 
and 17 to ground ($|Z_{1}|$, $|Z_{13}|$ and $|Z_{17}|$) versus the frequency of input circuit are shown in 
Fig. \ref{Fig.3} (c) and Fig. \ref{Fig.4} (c) for trimer and tetramer chain, respectively. The peak frequencies 
of the impedances are in good agreement with the calculated eigenvalues, despite some slight frequency 
shift of the measured impedance peaks due to component tolerances. In Fig. \ref{Fig.3} (c), the highest
impedance peak near 9.57 MHz of edge node inside the band gap (about 9.25 - 10.1 MHz) with 
impedance valleys for bulk nodes denotes the topological modes unambiguously. More intuitively, We 
measure the impedance distribution of the degenerate topological edge modes with strong locality at 
both ends at 9.57 MHz shown in Fig. \ref{Fig.3} (d). Differently, in Fig. \ref{Fig.4} (c), the impedance peak 
of edge node is accompanied by the impedance peaks of the bulk nodes near the frequency of edge state 
10.52 MHz, representing the existence of a topological bound state in a nontopological continuum where 
the bound edge state are show in Fig. \ref{Fig.4} (d).

In summary, we theoretically and experimentally demonstrated degenerate topological edge states in a 
class of topological multimer chains consisting of identical resonators with uniform hopping. By designing 
deliberate coupling paths, the systems exhibit full multimerization with fully dimerized SSH subspaces 
corresponding to flat bands that can be separated from their Hilbert spaces. We also show topological 
BICs by embedding the degenerate topological bound states into a continuous band in even chains 
naturally. Our scheme is experimental accessiable and can also be implemented in coupled waveguides 
arrays \cite{PhysRevLett.127.147401,Yan:23}, optical and acoustic coupled cavity arrays 
\cite{PhysRevResearch.3.013122, PhysRevLett.118.166803}, cold atoms lattices \cite{Goldman_2014} 
and three-dimensional circuit quantum electrodynamics \cite{PhysRevLett.124.023603}. Our work 
sheds new light on the construction of topological phases.

This work is supported by National Key Research and Development Program of China (2021YFA1400600, 
2021YFA1400602), the NSERC Discovery Grants and NSERC Discovery Accelerator Supplements (C.-M. H.), 
the National Natural Science Foundation of China (12274326) and China Scholarship Council 
(202106260079).

\bibliography{DTES.bib}

\end{document}